\documentclass[reprint,amsmath,amssymb,aps,prl,superscriptaddress]{revtex4-1}

\usepackage{graphicx}
\usepackage{dcolumn}
\usepackage{amsmath}
\usepackage{bm}
\usepackage[utf8]{inputenc}
\usepackage{mathtools}
\usepackage{pgffor}
\usepackage[normalem]{ulem}
\usepackage{color}
\usepackage{pdfpages}
\usepackage{BOONDOX-cal}

\makeatletter

\AtBeginDocument{\let\LS@rot\@undefined}
\makeatother

\newif\ifarXiv
\arXivtrue

\begin{document}

\title{Hydrodynamic and ballistic transport over large length scales in GaAs/AlGaAs}

\author{Adbhut Gupta}
\author{J. J. Heremans}
\email{heremans@vt.edu}
\affiliation{ Department of Physics, Virginia Tech, Blacksburg, Virginia 24061, USA}
\author{Gitansh Kataria}
\author{Mani Chandra}\email{manic@illinois.edu}
\affiliation{Research Division, Quazar Technologies, Sarvapriya Vihar, New Delhi 110016, India}
\author{S. Fallahi}
\affiliation{Department of Physics and Astronomy, Purdue University, West Lafayette, Indiana 47907, USA}
 \affiliation{Birck Nanotechnology Center, Purdue University, West Lafayette, Indiana 47907, USA} 
\author{G. C. Gardner}
 \affiliation{Birck Nanotechnology Center, Purdue University, West Lafayette, Indiana 47907, USA} 
 \affiliation{Microsoft Quantum Purdue, Purdue University, West Lafayette, Indiana 47907, USA}
\author{M. J. Manfra}
\affiliation{Department of Physics and Astronomy, Purdue University, West Lafayette, Indiana 47907, USA}
\affiliation{Birck Nanotechnology Center, Purdue University, West Lafayette, Indiana 47907, USA} 
\affiliation{Microsoft Quantum Purdue, Purdue University, West Lafayette, Indiana 47907, USA}
\affiliation{School of Electrical and Computer Engineering, Purdue University, West Lafayette, Indiana 47907, USA} 
\affiliation{School of Materials Engineering, Purdue University, West Lafayette, Indiana 47907, USA} 

\begin{abstract}
We study hydrodynamic and ballistic transport regimes through nonlocal resistance measurements and high-resolution kinetic simulations in a mesoscopic structure on a high-mobility two-dimensional electron system in a GaAs/AlGaAs heterostructure. We evince the existence of collective transport phenomena in both regimes and demonstrate that negative nonlocal resistances and current vortices are not exclusive to only the hydrodynamic regime. The combined experiments and simulations highlight the importance of device design, measurement schemes and one-to-one modeling of experimental devices to demarcate various transport regimes.
\end{abstract}
\maketitle

Electron transport in metals is often governed by momentum dissipation from electrons to the lattice, e.g., via impurity or phonon scattering. Such diffusive transport occurs when the momentum-relaxing (MR) electron mean-free path $\mathcal{l_\textrm{MR}}$ (obtained from electron mobility) is the shortest length scale in the system. However, in ultraclean two-dimensional electron systems (2DESs), a departure from diffusive transport occurs due to a long $\mathcal{l_\textrm{MR}}$, giving rise to either ballistic or hydrodynamic transport \cite{deJong}. In the ballistic regime, scattering mainly arises at the device boundaries, specularly or diffusively, and is delineated by the device scale $W$ \cite{heremans1992observation,heremans1994}. Yet inelastic electron-electron (\emph{e-e}) interactions transfer momentum predominantly among the electrons instead of to the lattice, conserving momentum within the electron system. When such momentum-conserving (MC) scattering$-$characterized by MC scattering mean-free path $\mathcal{l_\textrm{MC}}-$dominates, electrons can move collectively like a fluid and exhibit several effects associated with fluid dynamics \cite{Gurzhi, deJong, govorov2004, torre2015nonlocal, pellegrino2016, levitov2016electron, guo2017higher, chandra2019, shytov2018, nagaev2020, ledwith2019}. The observation of this hydrodynamic regime in electronic systems has attracted significant interest \cite{taubert2010, taubert2011, moll2016, gusev2018, levin2018, Braem2018, bandurin2016, kumar2017, bandurin2018, berdyugin2019, galitski2018, svintsov2018, gooth2018, kim2020, ku2020, raichev2020}.

The hydrodynamic regime shows a nonlocal current-voltage relation in devices, which can result in a negative nonlocal resistance ($R_\textrm{nl}$) \cite{chandra2019, torre2015nonlocal, pellegrino2016, levitov2016electron, guo2017higher}. Such sign reversal has been exploited in recent experiments to detect the onset of the hydrodynamic regime \cite{levin2018, bandurin2016, kumar2017, bandurin2018, berdyugin2019}.
However, the ballistic regime \emph{also} shows a nonlocal current-voltage relation, and can likewise produce negative $R_\textrm{nl}$ \cite{shytov2018, chandra2019}. In fact, the ballistic regime also supports striking current vortices and collective motion of particles usually associated with fluid depictions \cite{chandraquantum2019}. In this Letter, we reveal notable current vortices in both hydrodynamic and ballistic regimes, uniquely supported by evidence from elaborate measurements of $R_\textrm{nl}$. The presence of vortices in both hydrodynamic and ballistic regimes can be traced to electron momentum conservation in both regimes \cite{chandra2019, chandraquantum2019}.

\begin{figure*}[!t]
\begin{center}
\includegraphics[width=7 in]{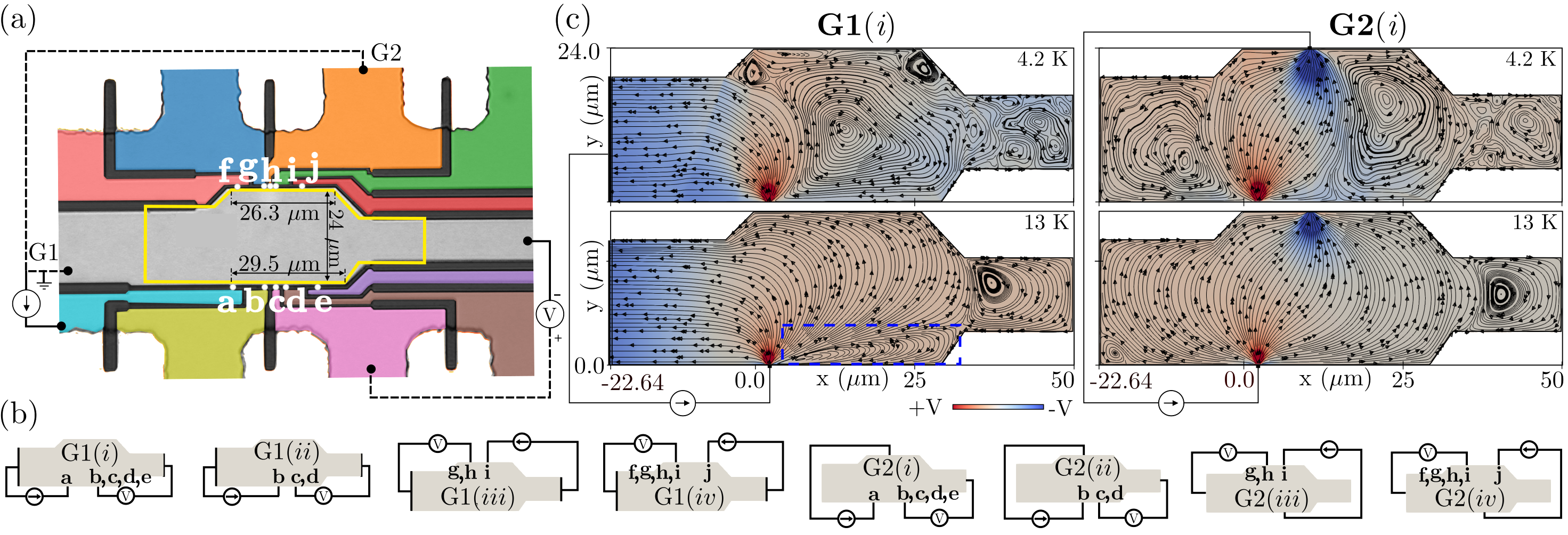}
\caption{\label{fig:fig1}  (a) Optical micrograph of the geometry showing dimensions, PCs $a-j$ (indicated by white dots, with paths to PCs in distinct colors), the computational domain (yellow outline) and current and voltage PCs for example measurements in the G1($i$) and G2($i$) configurations. Unused PCs are left floating. (b) Schematics of subconfigurations G1($i-iv$) and G2($i-iv$) depending on current injection PC and voltage measurement PC. (c) Simulated current streamline and voltage contour plots  for G1($i$) and G2($i$) at $T$ = 4.2 K and 13 K. The streamline and source arrows depict the direction of conventional current.}
\end{center}
\end{figure*}

We present strides in experimental device design, measurement schemes, and concomitant results, as well as in one-to-one modeling of the experimental device, pivotal in demarcating the transport regimes. We demonstrate measurements of $R_\textrm{nl}$ in a large-scale ($\sim 30\times24$ $\mu$m) ultraclean ($\mathcal{l_\textrm{MR}}\simeq 65$ $\mu$m at 4.2 K) device, which by its scale offers exceptional sensitivity to MC scattering, and hosts 10 point contacts (PCs) to probe voltages at various distances $\Delta x$ between the current injection point and voltage probes [Figs.~\ref{fig:fig1}(a) and \ref{fig:fig1}(b)]. The measurements at various $\Delta x$ are critical to check against the predictions of ballistic or hydrodynamic models. The exceptionally long $\mathcal{l_\textrm{MR}}$, due to optimized GaAs/AlGaAs MBE growth, favors the appearance of nondiffusive transport regimes. We interpret the experimental results using realistic high-resolution simulations of quasiparticle transport at the kinetic level, involving the actual experimental geometry in the precise contact configuration, and taking into account both MR and MC scattering. The simulations with $\mathcal{l_\textrm{MR}}$ and $\mathcal{l_\textrm{MC}}$ as inputs, determine that the device transitions from a predominantly ballistic regime at $T=4.2$ K to a hydrodynamic regime at $T\approx 12-19$ K [Fig.~\ref{fig:fig1}(c)].

Mesoscopic geometries were patterned on a GaAs/AlGaAs heterostructure containing a 2DES with mobility $\mu$ exceeding 670 m$^2$/(V s) at 4.2 K. The areal electron density is $N_{S} \approx$ 3.4$\times$10$^{15}$ m$^{-2}$, corresponding to a Fermi energy $E_{F} \approx$ 11.2 meV and $\mathcal{l_\textrm{MR}}$ = 64.5 $\mu$m at 4.2 K (Supplemental Material Sec. 1 \cite{Supp}). To measure $R_\textrm{nl}$, we fabricated an in-line mesoscopic geometry [Fig.~\ref{fig:fig1}(a)] containing 10 PCs ($a-j$) located in barriers which are on both sides of a multiterminal Hall mesa, with sides separated by $W \approx$ 24 $\mu$m. Each PC can act as either a current injector $\alpha$ (injecting electrons) or a voltage detector $\beta$ (detecting a nonlocal voltage). A measurement is defined by a pair consisting of an injector PC $\alpha$ and a detector PC $\beta$ in the same barrier, and we label the center-to-center separation between PCs in such pair as $L_{\alpha\beta}$, where $L_{\alpha\beta}$ ranges from 1.3 $\mu$m to 20.5 $\mu$m. Calling $V_\textrm{nl}$ the nonlocal voltage measured at $\beta$ vs a faraway counterprobe if current $I$ is injected at $\alpha$ and drained at another faraway counterprobe [Fig.~\ref{fig:fig1}(a)], the four-probe nonlocal resistance is expressed as $R_\textrm{nl}=V_\textrm{nl}/I$ and $R_\textrm{nl}$ takes the sign of $V_\textrm{nl}$. Measurements were performed in the linear response regime i.e. for small excitation energies (such that the system is everywhere close to equilibrium, with electrons close to $E_F$) over 4.2 K $\leq$ $T$ $\leq$ 40 K, using low-frequency ($\sim$ 44 Hz) ac lock-in techniques without dc offsets. We use $I\sim200$ nA, which is large enough for formation of vortices (Supplemental Material Sec. 4 \cite{Supp}) yet small enough to avoid electron heating. The conducting PC width $w \approx 0.6\, \mu$m and the Fermi wavelength $\lambda_F =$ 43 nm show that $w/(\lambda_F /2) \approx$ 28 spin-degenerate transverse modes contribute to transport, yielding a PC resistance $\approx (h/2e^2)/28$ = 461 $\Omega$ (Supplemental Material Sec. 1 \cite{Supp}). The large number of modes implies that PCs are very much open and act as classical PCs. The barriers and boundaries were defined using wet etching, which results in predominantly specular boundary scattering \cite{heremans1999, chen2005}. We exploit the flexibility provided by the geometry, allowing testing of different configurations for current injector and drain, and for many $\Delta x$, in the \emph{same} device. We use two current configurations: G1 where after injection at $\alpha$, $I$ is drained at the side of the device, and G2 where $I$ is drained at a PC at the opposite side of the device [Figs.~\ref{fig:fig1}(a) and \ref{fig:fig1}(b)]. The sensitivity to MC scattering turns out much higher in G1 (vicinity geometry \cite{bandurin2016}) than in G2 (Supplemental Material Sec. 7 \cite{Supp}).  

\begin{figure}[!b]
\includegraphics[width=1\linewidth]{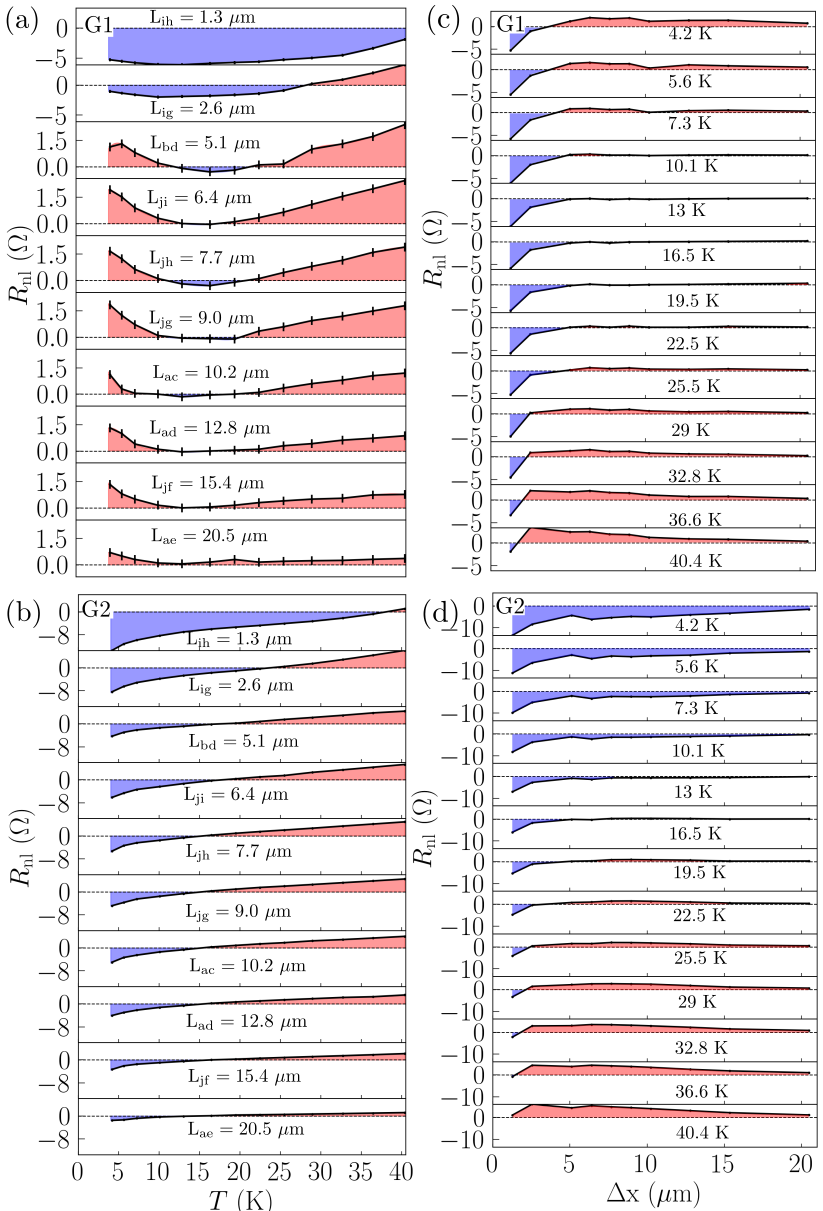}
\caption{(a),(b) Experimental nonlocal resistance $R_\textrm{nl}$ vs $T$ for each $L_{\alpha\beta}$ for (a) G1 and (b) G2. (c),(d) $R_\textrm{nl}$ vs $\Delta x$  for each $T$ for (c) G1 and (d) G2. The dotted lines indicate $R_\textrm{nl} = 0$, with negative (positive) regions of $R_\textrm{nl}$ shaded in blue (red).}
\label{fig:fig2}
\end{figure}

Transport in the device is modeled via the Boltzmann equation \cite{chandra2019,Supp}
\begin{equation}
\frac{1}{v_F}\frac{\partial f}{\partial t}  + \left(\frac{\mathbf{p}}{mv_F}\right).\frac{\partial f}{\partial \mathbf{x}}  = -\frac{f - f_0^\textrm{MR}}{\mathcal{l_\textrm{MR}}} - \frac{f - f_0^\textrm{MC}}{\mathcal{l_\textrm{MC}}},
\label{eq:1}
\end{equation}
where $f(\mathbf{x}, \mathbf{p}, t)$ is the electron distribution in the spatial coordinates $\mathbf{x} \equiv (x, y)$, momentum coordinates $\mathbf{p} \equiv (p_x, p_y)$, and time $t$. While long-range electric fields are not explicitly present in Eq.~(\ref{eq:1}), they are included at linear order as the gradient of the electrochemical potential \cite{chandra2019}. The left side (with $v_F$ as the Fermi velocity and $m$ as the effective mass) describes free advection, and the right side thermalization due to MR and MC scattering in a relaxation time approximation \cite{Footnote} with $f_0^\textrm{MR}$ and $f_0^\textrm{MC}$ as the local stationary and drifting Fermi-Dirac distributions (details of the model can be found in \cite{chandra2019} and in Supplemental Material Sec. 2 \cite{Supp}). The model inputs are $\mathcal{l_\textrm{MC}}$ (a free parameter) and $\mathcal{l_\textrm{MR}}$ (fixed by $\mu$). We consider dynamics at the Fermi surface without thermal smearing so that $\mathbf{p} = m v_F \hat{p}$ and solve for transport in the zero-frequency limit  ($\partial/\partial t \rightarrow 0$); $v_F$ then factors out, leaving the circular Fermi contour as the only relevant detail. We solve Eq.~(\ref{eq:1}) in the precise experimental geometry using \textsc{bolt} \cite{bolt}, a high-resolution solver for kinetic theories. The overall prefactor of the numerical solutions is set by calibrating against the measurements in G1($ii$) [Fig.~\ref{fig:fig1}(b)] for each $T$ (Supplemental Material Sec. 3 \cite{Supp}). 

The experimental $R_\textrm{nl}$ vs $T$ for G1 and G2 are depicted in Figs.~\ref{fig:fig2}(a) and \ref{fig:fig2}(b) respectively, for the specific $L_{\alpha\beta}$ used in measurements [Fig.~\ref{fig:fig1}(a) exemplifies $L_{ac}$]. Two inferences appear: the negative $R_\textrm{nl}$ attest to a departure from diffusive transport, and a striking contrast exists in $T$ dependences between G1 and G2. For G1, $R_\textrm{nl}$ shows a nonmonotonic dependence on $T$, initially decreasing as $T$ increases, crossing over to negative values in a particular range of $T$ for given $L_{\alpha\beta}$, then increasing to positive values. For G2, $R_\textrm{nl}$ increases from negative values at low $T$ to positive values at higher $T$. The difference in $T$ dependence between G1 and G2 indicates that the current injector-drain configuration significantly affects transport. Figure~\ref{fig:fig2}(a) can be directly compared with similar results in graphene \cite{bandurin2016, bandurin2018} and other GaAs/AlGaAs experiments \cite{levin2018}. Figures~\ref{fig:fig2}(c) and \ref{fig:fig2}(d) depict $R_\textrm{nl}$ vs $\Delta x$ parametrized in $T$ for G1 and G2 respectively, tracing a crossover from negative to positive values vs $\Delta x$.

\begin{figure*}[!htbp]
\begin{center}
\includegraphics[width=7 in]{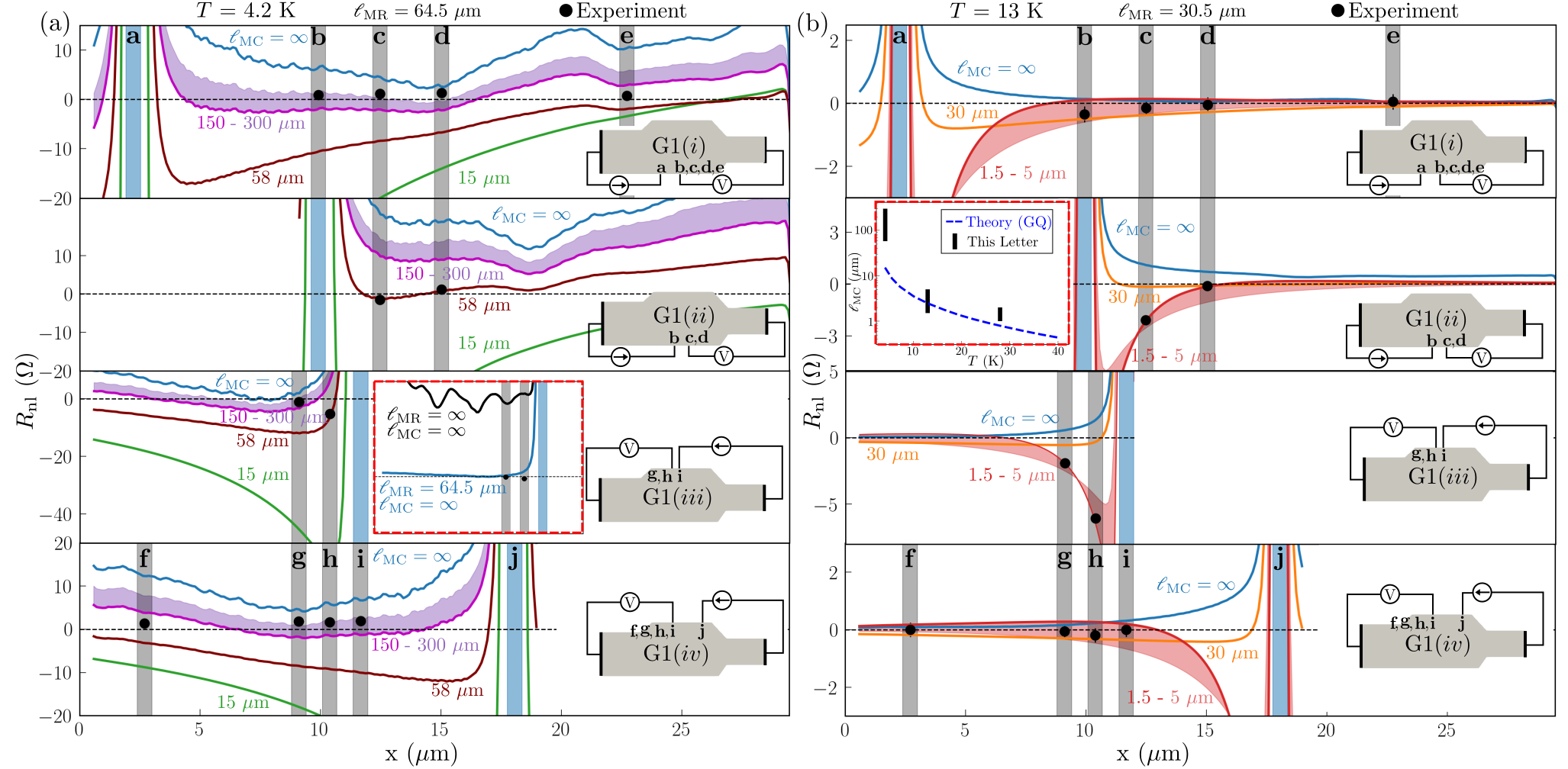}
\caption{Experimental $R_\textrm{nl}$ (black dots) and modeled $R_\textrm{nl}$ (lines) for G1($i-iv$), plotted vs location $x$ along the barrier into which the injection PC (blue vertical bars) is placed, at (a) $T=4.2$ K where $\mathcal{l_\textrm{MR}}$ = 64.5 $\mu$m and (b) $T=13$ K where $\mathcal{l_\textrm{MR}}$ = 30.5 $\mu$m. The modeled $R_\textrm{nl}$ (lines) is shown parametrized in $\mathcal{l_\textrm{MC}}$. Schematics of subconfigurations are also depicted. The grey vertical bars represent locations of detector PCs. Experimental $R_\textrm{nl}$ (black dots) for G1($ii$) are chosen for reference calibration (Supplemental Material Sec. 3 \cite{Supp}) given the clear crossover in $R_\textrm{nl}$ in G1($ii$). Inset in (a)(panel 3) shows $R_\textrm{nl}$ for G1($iii$) for ($\mathcal{l_\textrm{MC}}\rightarrow\infty, \mathcal{l_\textrm{MR}}\rightarrow\infty$) and for ($\mathcal{l_\textrm{MC}}\rightarrow\infty, \mathcal{l_\textrm{MR}}=$ 64.5 $\mu$m). Inset in (b)(panel 2) shows the extracted $\mathcal{l_\textrm{MC}}$ vs $T$ (black bars), along with theoretical estimates (blue dotted line) from GQ \cite{GQ1982} (Eq. S1 in Supplemental Material Sec. 5 \cite{Supp}).}
\label{fig:fig3}
\end{center}
\end{figure*}

In Fig.~\ref{fig:fig2}(a) for $T\lesssim 10$ K, G1 shows negative $R_\textrm{nl}$ for small $\Delta x \lesssim 2.6\,\mu$m ($L_{ih}, L_{ig}$), crossing over to positive $R_\textrm{nl}$ for $\Delta x\gtrsim 5.1\,\mu$m ($L_{bd}, L_{ji}$, ...). As $T$ is increased, $R_\textrm{nl}$ vs $T$ develops a minimum for all $\Delta x$, and the $R_\textrm{nl}$ at this minimum crosses over to negative values for $\Delta x \lesssim 12.8 \ \mu$m. These observations [confirmed in Fig.~\ref{fig:fig2}(c)] are a consequence of the interplay between $\mathcal{l_\textrm{MC}}(T)$, $\mathcal{l_\textrm{MR}}(T)$ and geometry. Considering only $\mathcal{l_\textrm{MR}}(T)$, with $\mathcal{l_\textrm{MR}}(4.2\mathrm{K})=64.5 \ \mu$m $\gg W = 24 \ \mu$m at $T=$ 4.2 K the system is predominantly ballistic. By finding the values of $\mathcal{l_\textrm{MC}}$ (used as a model input parameter, given $\mathcal{l_\textrm{MR}}$) for which the experimental and modeled $R_\textrm{nl}$ match, we have a means of bracketing values for $\mathcal{l_\textrm{MC}}(T)$. The procedure is illustrated in Fig.~\ref{fig:fig3}. In Fig.~\ref{fig:fig3}(a) we focus on $T$ = 4.2 K, lying in the region $T\lesssim 10$ K showing a crossover vs $\Delta x$ from negative $R_\textrm{nl}$ at small $\Delta x$ (1.3 and 2.6 \ $\mu$m) to positive further away [see Figs.~\ref{fig:fig2}(a) and \ref{fig:fig2}(c)]. The inset in Fig.~\ref{fig:fig3}(a)(panel 3) shows that the limiting billiard model, common in ballistic transport and using ($\mathcal{l_\textrm{MC}}\rightarrow\infty, \mathcal{l_\textrm{MR}}\rightarrow\infty$), does not capture the crossover vs $\Delta x$ because it results in $R_\textrm{nl}>0$ for all $\Delta x$ (inset, black trace); yet positive $R_\textrm{nl}$ is not universal in the billiard model and can be heavily influenced by geometry (Supplemental Material Sec. 6 \cite{Supp}). Introducing finite MR scattering via the experimental $\mathcal{l_\textrm{MR}}$ = 64.5 $\mu$m ($T$ = 4.2 K) and zero MC scattering with $\mathcal{l_\textrm{MC}}\rightarrow \infty$, the modeled $R_\textrm{nl}$ are lower compared to the billiard model but \emph{still} do not reach $R_\textrm{nl}<0$ at small $\Delta x$ (inset, blue trace). As shown in Fig.~\ref{fig:fig3}(a), only with finite MC scattering using a range $\mathcal{l_\textrm{MC}} \simeq 60-300$ $\mu$m ($\gg W = 24$ $\mu$m), does the model yield a crossover from negative to positive $R_\textrm{nl}$ with increasing $\Delta x$. The inset in Fig.~\ref{fig:fig3}(b)(panel 2) plots this range at 4.2 K as compatible with data and model. 

A theoretical prediction for $\mathcal{l_\textrm{MC}}$ is found in a commonly used theoretical expression for quantum lifetime by Giuliani and Quinn (GQ) \cite{GQ1982} [Eq. S1 in Supplemental Material Sec. 5 \cite{Supp}], also plotted in the inset. The inset demonstrates that $\mathcal{l_\textrm{MC}} \simeq 60-300$ $\mu$m exceeds values from GQ which at 4.2 K yields $\mathcal{l_\textrm{MC}} \approx 15$ $\mu$m. In fact, using $\mathcal{l_\textrm{MC}} = 15 \ \mu$m in the model yields $R_\textrm{nl}<0$ throughout the device [Fig.~\ref{fig:fig3}(a), green curve], contrary to experimental observations. Recent results, in fact, suggest that longer $\mathcal{l_\textrm{MC}}$ can result from dielectric screening \cite{kim2020}. In short, considering $\mathcal{l_\textrm{MC}}$, for $T\lesssim 10$ K neither the billiard model ($\mathcal{l_\textrm{MC}}\rightarrow\infty$) nor GQ (too short $\mathcal{l_\textrm{MC}}$) reproduce the experiments, and intermediate $\mathcal{l_\textrm{MC}}$ is required. Further, the presence of numerous current vortices of various sizes in the predominantly ballistic regime [Fig.~\ref{fig:fig1}(c) at 4.2 K], reveals that  dominance of MC scattering is not obligatory for the formation of vortices.

For $10 \ \mathrm{K} \lesssim T \lesssim 22.5 \ \mathrm{K}$ in G1, $R_\textrm{nl}$ vs $T$ develops a minimum, which becomes shallower with increasing $\Delta x$, and $R_\textrm{nl}$ crosses over to negative values for $\Delta x \lesssim 12.8 \ \mu$m [Fig.~\ref{fig:fig2}(a)]. Modeling using $\mathcal{l_\textrm{MC}} \simeq 1.5-5 \ \mu$m accommodates all the experimental data at $T=13$ K [Fig.~\ref{fig:fig3}(b)]. Since $\mathcal{l_\textrm{MC}}\ll W$ and $\mathcal{l_\textrm{MC}}\ll \mathcal{l_\textrm{MR}}$ while $\mathcal{l_\textrm{MR}}=30.5\,\mu$m $\gtrsim W$ is sufficiently long, the system is in the hydrodynamic regime. At $T$ = 13 K, GQ yields $\mathcal{l_\textrm{MC}} \simeq 2.5 \ \mu$m \cite{GQ1982}, lying in the model's range of $\mathcal{l_\textrm{MC}} \simeq 1.5-5 \ \mu$m. This suggests that using conditions where $\mathcal{l_\textrm{MC}} \gtrsim W$ (as at 4.2 K) leads to a more discriminating test of microscopic theories of $\mathcal{l_\textrm{MC}}$. As expected in the hydrodynamic regime, Fig.~\ref{fig:fig1}(c) (13 K) exhibits current vortices. 

Yet the current vortices in the hydrodynamic regime in Fig.~\ref{fig:fig1}(c) (13 K) exhibit a distinct pattern compared to the ballistic vortices at 4.2 K. In the hydrodynamic regime only a single vortex [dashed box in Fig.~\ref{fig:fig1}(c)] inhabits the main chamber, and this result is obtainable from just the Navier-Stokes fluid equations. In contrast, the ballistic profile [top panels in Fig.~\ref{fig:fig1}(c)] shows multiple vortices, which cannot be accessed from fluid equations and require solving a Boltzmann kinetic equation. Notably, while the fluid equations cannot access the ballistic limit, the Boltzmann equation can access the fluid solutions in the limit $\mathcal{l_\textrm{MC}} \ll W$.

As $T$ is increased further in G1 to $T\gtrsim$ 19 K, Fig.~\ref{fig:fig2}(a) shows an upward trend in $R_\textrm{nl}$ vs $T$ toward $R_\textrm{nl}>0$ for all $\Delta x$. The crossover to $R_\textrm{nl}>0$ occurs at higher $T$ for smaller $\Delta x$ ($R_\textrm{nl}$ at $\Delta x=$ 1.3 $\mu$m is still negative at 40 K but is estimated to go positive around 45 K). This is corroborated by Fig.~\ref{fig:fig2}(c) where the region of $R_\textrm{nl}<0$ diminishes to smaller $\Delta x$ with increasing $T$. This behavior heralds a breakdown of hydrodynamic transport and a transition from hydrodynamic to diffusive dynamics as MR scattering increasingly affects transport. In G1 the hydrodynamic regime exists at intermediate $T\approx 12-19$ K, estimated from Fig.~\ref{fig:fig2}(c) by tracing the crossover from negative to positive $R_\textrm{nl}$ vs  $\Delta x$. Supplemental Material Sec. 8 \cite{Supp} shows the experimental and modeled $R_\textrm{nl}$ at 28 K, and the change in vortex pattern with changing $\mathcal{l_\textrm{MC}}$.

Figures~\ref{fig:fig2}(b) and \ref{fig:fig2}(d) reveal that in the G2 configuration $R_\textrm{nl}<0$ occurs over a wide range of $T$, rendering it inefficient in differentiating between ballistic and hydrodynamic regimes. The simulations also show that G2 is \emph{insensitive} to MC scattering (Supplemental Material Sec. 7 \cite{Supp}). Both properties disallow using G2 to determine $\mathcal{l_\textrm{MC}}$, highlighting the importance of choosing appropriate contact configurations for discriminating transport regimes.

In conclusion, nondiffusive transport, either predominantly ballistic or hydrodynamic, is realized over a wide temperature range in a large-scale GaAs/AlGaAs geometry. The appearance of both predominantly ballistic or hydrodynamic regimes at such a large scale, despite opposite required limits of the strength of MC scattering, is striking. Equally remarkable are their shared characteristics of negative nonlocal resistances and current vortices. The nonlocal resistance in both regimes can be tuned by device and contact geometry, used here to disentangle the regimes and to obtain a measure of the MC scattering length. While the importance of geometry in the ballistic regime is well known, we additionally find that the ballistic regime can also exhibit collective effects, such as current vortices, even in the absence of dominant electron-electron interactions in a large-scale ultraclean device.

The authors acknowledge support by the U.S. Department of Energy, Office of Basic Energy Sciences, Division of Materials Sciences and Engineering under awards DE-FG02-08ER46532 (J. J. H.) and DE-SC0020138 (M. J. M.). The authors acknowledge computational resources (GPU clusters {\tt cascades} and {\tt newriver}) and technical support provided by Advanced Research Computing at Virginia Tech. We thank Ravishankar Sundararaman for helpful discussions. M. C. thanks Sujit Kumar for hospitality during the course of this work.



\section*{Supplementary material}

# Supplemental Material to "Hydrodynamic and ballistic transport over large length scales in GaAs/AlGaAs"

Adbhut Gupta,[1] J. J. Heremans,[2] Gitansh Kataria,[3] Mani Chandra,[3]
S. Fallahi,[4,5] G. C. Gardner,[5,6] and M. J. Manfra[4,5,6,7,8]

[1]*Department of Physics, Virginia Tech, Blacksburg, VA 24061, USA*
[2]*Department of Physics, Virginia Tech, Blacksburg, VA, USA*
[3]*Research Division, Quazar Technologies, Sarvapriya Vihar, New Delhi 110016, India*
[4]*Department of Physics and Astronomy, Purdue University, West Lafayette, IN 47907, USA*
[5]*Birck Nanotechnology Center, Purdue University, West Lafayette, IN 47907, USA*
[6]*Microsoft Quantum Purdue, Purdue University, West Lafayette, IN 47907, USA*
[7]*School of Electrical and Computer Engineering,
Purdue University, West Lafayette, IN 47907, USA*
[8]*School of Materials Engineering, Purdue University, West Lafayette, IN 47907, USA*

## DEVICE FABRICATION AND MATERIALS PROPERTIES

The devices were fabricated from GaAs/AlGaAs MBE-grown material (Fig. 1(a) main text). The GaAs quantum well hosting the two-dimensional electron system (2DES) is located 190 nm below the surface, has a width of 26 nm, and is top- and bottom-doped by Si $\delta-$layers 80 nm removed from the quantum well and embedded in $Al_{0.32}Ga_{0.68}As$ barriers. Optimization of heterostructure design is described in Ref. [1]. The mesoscopic geometries were patterned by electron beam lithography and wet etching of the barriers, using PMMA as the etching mask. The geometries feature mesoscopic apertures (point contacts, PCs) separated by various distances $L_{\alpha\beta}$ on both sides of a Hall mesa. Different $\alpha$ (current injector) and $\beta$ (voltage detector) are chosen such that a wide range of distances $L_{\alpha\beta}$ exists, from 1.3 $\mu$m to 20.5 $\mu$m. The PC resistance $R_{pc}$ varies between 450 $\Omega$ to 750 $\Omega$ at $T = 4.2$ K, depending on the PC.

The electron transport properties of the unpatterned 2DES were characterized by the van der Pauw method. We use the value of 2D resistivity $R_\square$ from this method, and areal electron density $N_S$ from Hall measurements on the fabricated device to obtain electron mobility $\mu$. At temperature $T = 4.2$ K, we obtain $N_S = 3.4 \times 10^{15}$ m$^{-2}$ and $R_\square = 2.74$ $\Omega/\square$, yielding $\mu = 670$ m$^2$/Vs (confirming the quality of the material), Drude (mobility) mean-free-path $\ell_{mr} = 64.5$ $\mu$m and Fermi energy $E_F = 11.2$ meV (equivalent to $\sim 130$ K). Here, $\ell_{mr} = v_F \tau_{mr}$, with $v_F$ the Fermi velocity and $\tau_{mr}$ the Drude momentum relaxation time derived from $\mu = e\tau_{mr}/m$, where $m$ denotes the electron effective mass (0.067 $m_e$ with $m_e$ the free-electron mass) and $e$ the electron charge. The momentum-relaxing (MR) electron mean-free-path $\ell_{mr}$ describes momentum dissipation from electrons to the lattice e.g. via impurity or phonon scattering. In contrast the momentum-conserving (MC) mean-free-path $\ell_{mc}$ describes transfer of momentum internal to the 2DES between electrons via electron-electron scattering, conserving momentum within the 2DES. Since MC scattering merely causes a redistribution of momentum internally to the electron fluid, and since $\mu$ quantifies the loss of total momentum of the electron fluid to the lattice, $\mu$ cannot provide a measure of $\ell_{mc}$.

Non-parabolicity of the band structure is accounted for in calculating the transport properties [2, 3]. $N_S$ (Fig. S1(a)) and $R_\square$ (Fig. S1(b)) are observed to increase with increasing $T$, while $\mu \sim 1/T$ (Fig. S1(c)), demonstrating that as expected $\mu$ is limited by scattering with acoustic phonons. However we observe a slight change in slope of $\mu$ vs $T$ for $T \approx 13$ K (inset in Fig. S1(c)), attributed to incipient scattering by LO phonons for $T > 13$ K. In a high-$\mu$ 2DES like the present, LO phonon scattering can start to be observed even at $T \approx 13$ K since the lack of residual scattering does not mask their effect, while in a 2DESs of lower $\mu$ the contribution to scattering by optical phonons is only apparent at higher $T$. As a result, the dependence on $T$ of $\mu$ is affected slightly. Figure S1(c) also shows that $\ell_{mr} \sim 1/T$, again demonstrating the expected dominance of acoustic phonons in limiting $\ell_{mr}$. The fact that $\ell_{mr}$ and $\mu$ follow closely analogous dependences on $T$, highlights that $\mu$ can provide a measure of $\ell_{mr}$ but not of $\ell_{mc}$. Figure S1(d) depicts $1/\mu$ vs $T$, indicating that a relation $1/\mu(T) = 1/\mu_o + \alpha T$ is closely followed, where $\mu_o$ denotes $\mu$ limited by impurity scattering and $\alpha T$ describes the linear dependence on $T$ due to (predominantly) acoustic phonons where $\alpha$ is a proportionality constant [4]. The exponent of $T$ changes from 1 to 1.2 at $T > 13$ K, but since the deviation due to optical phonons is small, the dependence due to acoustic phonons is a good approximation.

## MODELING APPROACH

In the Boltzmann equation (Eq. 1 main text and Ref. [5]), $f_0^{mr}$ and $f_0^{mc}$ denote local equilibrium distributions, to which an arbitrary distribution $f(\mathbf{x},\mathbf{p})$ relaxes by the action of MR and MC scattering respectively [5]. Math-

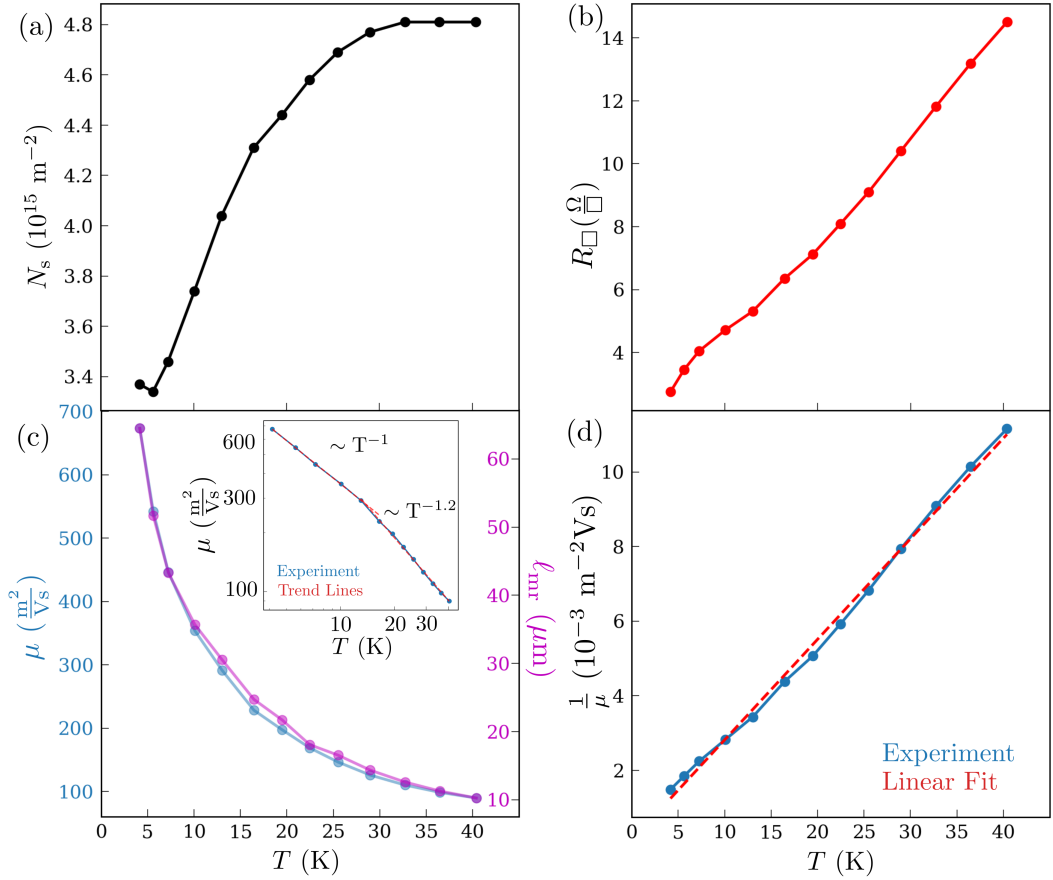

Fig. S1. (a) Areal carrier density $N_S$ vs $T$. (b) 2D resistivity $R_\Box$ vs $T$. (c) $\mu$ vs $T$ (left axis) and $\ell_{mr}$ vs $T$ (right axis), with inset depicting $\mu$ vs $T$ on log-log scale, highlighting the change in slope. (d) $1/\mu$ vs $T$. From the fit (red curve), we extract the fitting parameters $1/\mu_o = 1.2 \times 10^{-4}$ m$^{-2}$Vs and $\alpha = 2.7 \times 10^{-4}$ m$^{-2}$VsK$^{-1}$.

ematically, $f_0^{mr}(\mathbf{x}, \mathbf{p}) = f_0(\mu^{mr})$, where $f_0$ denotes the Fermi-Dirac distribution (in this work taken to be at $T = 0$). The variable $\mu^{mr}(\mathbf{x})$ is a spatially varying chemical potential, solved for by imposing conservation of local particle number $\int f d^2p = \int f_0^{mr} d^2p$ (the only local conserved property of MR scattering). For MC scattering, conservation of momentum necessitates an additional variable to describe the local equilibrium, given by $f_0^{mc} = f_0(\mu^{mc} + \mathbf{p} \cdot \mathbf{v}_d)$, where $\mathbf{v}_d(\mathbf{x})$ is a spatially varying drift velocity. The variables $\mu^{mc}$ and $\mathbf{v}_d$ are solved for by invoking the conservation properties of MC scattering: local number conservation $\int f d^2p = \int f_0^{mc} d^2p$ and local momentum conservation $\int \mathbf{p} f d^2p = \int \mathbf{p} f_0^{mc} d^2p$ [5].

## CALIBRATION BETWEEN EXPERIMENTS AND MODELING

The solution of the kinetic model (Eq. 1 main text) is the non-equilibrium distribution function $f(\mathbf{x}, \mathbf{p})$. To obtain voltages, we would further need to solve the Poisson equation sourced by the 2D density $N_S + \delta N_s(\mathbf{x}) \sim \int f(\mathbf{x}, \mathbf{p}) d^2p$, while taking into account the full 3D electrostatic environment. We forgo this complexity, and instead directly calibrate the spatial density variation, $\delta N_s(\mathbf{x})$ obtained from the model, to the measured nonlocal voltages $V_{nl}(\mathbf{x})$ in one configuration (chosen to be G1($ii$), but can be any of the four configurations (Fig. 1(b) main text)). The measurements yield the electrochemical potential at given $\mathbf{x}$. Since the density of states is constant over energy in a 2DES, the change in electrochemical potential at $\mathbf{x}$ is directly proportional to the density variation at $\mathbf{x}$. This results in the local capacitance approximation, where $V_{nl}(\mathbf{x}) \propto \delta N_s(\mathbf{x})$, with a constant prefactor determined by the electrostatic environment. We thus have nonlocal resistance $R_{nl}(\mathbf{x}) = A \delta N_s(\mathbf{x})$, where $A$ denotes the calibration prefactor. Since $|\delta N_s(\mathbf{x})| << N_s$ and hence $|eV_{nl}(\mathbf{x})| << E_F$, the measurements are performed in a linear regime. This also holds for locations near PCs since $|eR_{pc}I| << E_F$ indicating that close to the current injection PCs, the electrons have energies very close to $E_F$.

To perform the calibration, we need to find $(A, \ell_{mc})$ such that $R_{nl}$ obtained from simulations in G1($ii$) match

experimental values. Given freedom to set $A$, we find that a match occurs only for a specific value of $\ell_{mc}$, since the solution should agree with the measured $R_{nl}$ at *all* PCs. The allowed range can be efficiently bisected by first matching against the *sign* of $R_{nl}$ (cfr G1($ii$) in Fig. 3 main text). The calibration therefore results in a pair of unique $(A, \ell_{mc})$. After $A$ has been set, the range of $\ell_{mc}$ required for the modeling to match experiments in the other configurations represents an error in the method. Finally, we note that we had to perform the calibration procedure separately for each $T$, i.e. $A \equiv A(T)$. The reason for this dependence on $T$ is left for future work.

The low-frequency ($\sim 44$ Hz) AC lock-in measurement, performed without DC offset, tracks the signs of $\delta N_s(\mathbf{x})$ and $R_{nl}(\mathbf{x})$ such that for $\delta N_s(\mathbf{x}) > 0$ (overdensity of electrons at $\mathbf{x}$), we measure $R_{nl}(\mathbf{x}) > 0$ and for $\delta N_s(\mathbf{x}) < 0$ (underdensity of electrons at $\mathbf{x}$), we measure $R_{nl}(\mathbf{x}) < 0$.

## CONDITIONS FOR FORMATION OF VORTICES

For formation of vortices, we require that a sufficient number of quasiparticles participate in transport since otherwise the collective effects will be swamped out by shot noise. For an injected current $I$, the number of quasiparticle excitations in the device is given by $N = It_d/e$, where $t_d$ denotes the dwell time which each injected quasiparticle spends in the device before exiting through either source or drain. Vortices require $N \gg 1$, and thus require $I \gg e/t_d$. In the ballistic regime, $t_d = L_{\text{path}}/v_F$, where $L_{\text{path}}$ is the path length traversed by the quasiparticle before exiting the device. The path taken by ballistic quasiparticles from their injection point to their exit point can be tortuous: they experience several reflections off the device boundary before exiting, and hence the condition $L_{\text{path}} > W$, where $W$ is the device scale, is readily satisfied. We parametrize the geometric uncertainty into a small dimensionless parameter $\epsilon < 1$ and write $L_{\text{path}} = W/\epsilon$ such that $t_d = W/(\epsilon v_F)$. Thus, we finally obtain the vortex formation condition $I \gg \epsilon e(v_F/W)$. For our device over the range of $T$, $v_F = (2.32 - 2.72) \times 10^5$ m/s, yielding $W/v_F \approx 88$ ps - 103 ps. Hence we require $I \gg \epsilon$ (1.5 nA - 1.8 nA). Therefore, even assuming $\epsilon \lesssim 1$, the injection current in our experiment ($I = 200$ nA) is far above the threshold required for vortices to form.

## EFFECT OF DEVICE GEOMETRY AND CONTACT CONFIGURATION ON THE BALLISTIC REGIME

In the ballistic regime, the device boundaries play an important role, and in fact the ballistic regime can be tuned by choosing an appropriate geometry. The exper-

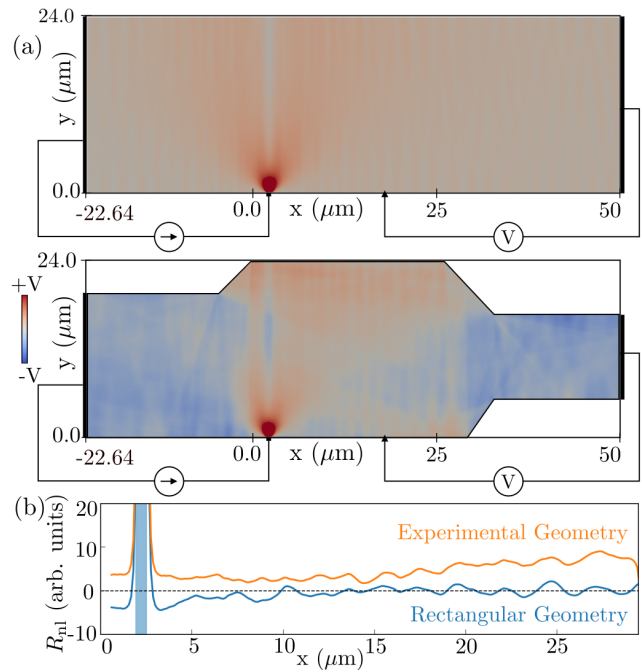

Fig. S2. (a) Modeled voltage contour plots in the simple rectangular (top) and experimental (bottom) geometry in the ballistic limit, $(\ell_{mc}, \ell_{mr}) = (\infty, \infty)$ in the G1($i$) configuration. (b) Modeled $R_{nl}$ vs $x$ for both geometries, where $x$ denotes the horizontal location along the lower barrier. The blue vertical bar represents the location of the injector PC.

imental geometry has a complex shape, with corners of the main chamber defined by slanting edges which taper down to the side ports. As we show below, this shape requires a more complicated numerical discretization scheme than needed for a simpler geometry such as a rectangle but helps us disentangle the ballistic and hydrodynamic regimes.

We emphasize here that modeling the precise experimental geometry is crucial to interpreting results in the ballistic regime. $R_{nl}$ can change *qualitatively* with changes in geometry. We illustrate this sensitivity by the G1($i$) configuration, with current injected from the lower barrier and extracted to the left in Fig. S2(a). We consider the ballistic limit with $(\ell_{mc}, \ell_{mr}) = (\infty, \infty)$ (billiard model). As depicted in Fig. S2(b), for the simple rectangular geometry in Fig. S2(a), $R_{nl} < 0$ for distances $\Delta x < 10 \, \mu$m between current injection and voltage detection PCs, beyond which $R_{nl}$ varies between positive and negative values. On the other hand, $R_{nl}$ in the experimental geometry in Fig. S2(a-b) remains positive throughout $\Delta x$. The reason for $R_{nl} > 0$ becomes clear on examining the carrier density distribution from the modeling: the slanting edges reflect and focus carriers back into the main chamber and increase the density relative to the grounded side ports, giving rise to $R_{nl} > 0$ throughout $\Delta x$.

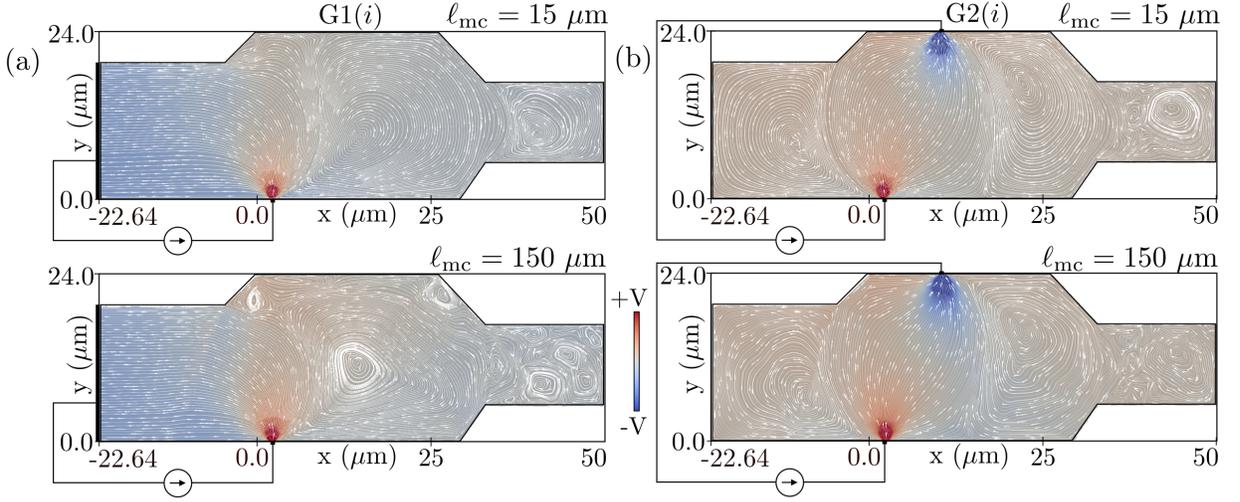

Fig. S3. Current streamline and voltage contour plots obtained from simulations for (a) G1($i$) and (b) G2($i$) as $\ell_{mc}$ is changed from 15 $\mu$m to 150 $\mu$m at $T = 4.2$ K where $\ell_{mr} = 64.5\,\mu$m. G1 ($i$) shows higher sensitivity to $\ell_{mc}$, with a single large current vortex at $\ell_{mc} = 15$ $\mu$m breaking down to multiple smaller vortices at $\ell_{mc} = 150$ $\mu$m. In G2 ($i$), the vortex pattern does not significantly change under even the 10× change in $\ell_{mc}$, indicating the insensitivity of the G2 configuration to $\ell_{mc}$.

The ballistic response also depends on the contact configuration, even within the same overall device geometry. On changing the contact configuration to G2, $R_{nl} < 0$ prevails at the location of almost all the PCs, becoming undifferentiated from the hydrodynamic regime (Fig.2(b) main text). Therefore, interpreting the experimental data requires taking into account the precise device geometry as well as the contact configuration.

## EFFECT OF CONTACT CONFIGURATION ON SENSITIVITY TO $\ell_{mc}$

For a fixed device geometry the contact configuration can strongly affect the sensitivity of the device to MC scattering (Fig. S3). We find that the G2 configuration is markedly *insensitive* to changes in $\ell_{mc}$, as illustrated in Fig. S3(b) by the modeled current streamline and voltage contour plots as $\ell_{mc}$ is varied. Furthermore, $R_{nl} < 0$ in G2 for both ballistic and hydrodynamic regimes. G2 is therefore ineffective to probe hydrodynamic transport. In contrast, G1 shows distinctly higher sensitivity to $\ell_{mc}$, as depicted in Fig. S3(a). For the given device geometry, G1 allows a differentiation between ballistic and hydrodynamic transport.

## EFFECT OF MR SCATTERING AT HIGHER $T$

In the experimental geometry, a surprising behavior is encountered in the interplay between MR and MC scattering in the presence of stronger MC scattering ($\ell_{mc} \lesssim W$) at higher $T$: as MC scattering increases with

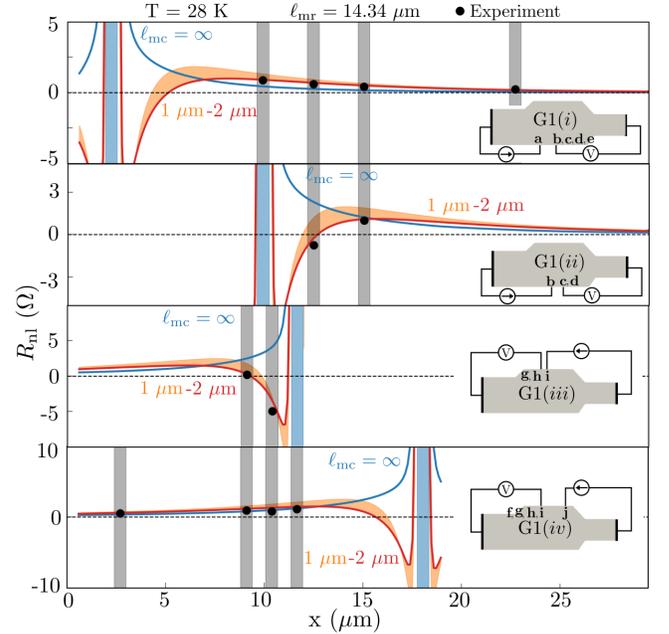

Fig. S4. Modeled $R_{nl}$ for G1 with $\ell_{mr} = 14.3$ $\mu$m (corresponding to $T = 28$ K) for $\ell_{mc} = 1 - 2$ $\mu$m and $\ell_{mc} \to \infty$, plotted vs location $x$ along the barrier into which the injection PC (blue vertical bar) is placed, along with the experimental data points (black dots). The grey vertical bars represent locations of detector PCs. At low $\ell_{mc}$ (higher $T$), $\ell_{mr}$ exerts a dominant effect on $R_{nl}$ for $\Delta x \gtrsim 3\,\mu$m, where the curves for $\ell_{mc} = 1 - 2$ $\mu$m and $\ell_{mc} \to \infty$ lie within $\sim 1$ Ohm. Determination of $\ell_{mc}$ from such measurements is difficult, unless measurements are chosen close to injector PC(G1(ii) and G1(iii)) where $R_{nl} \lesssim 0$.

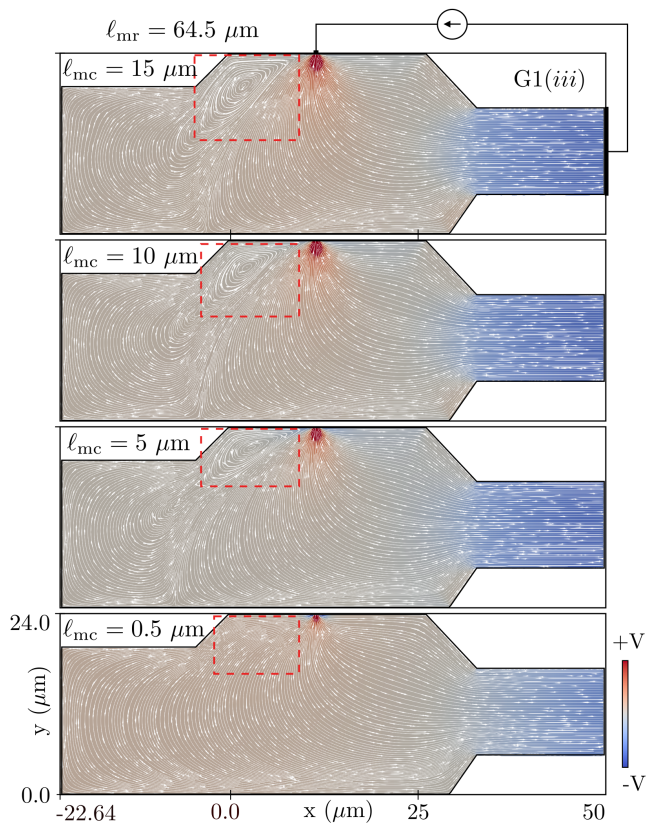

Fig. S5. Modeled current streamline and voltage contour plots for G1(*iii*) at fixed $\ell_{mr} = 64.5$ µm for $\ell_{mc}$ decreasing panel to panel, top to bottom. The sequence shows the disappearance of the vortex (dotted red box) with decreasing $\ell_{mc}$ for a finite $\ell_{mr}$.

increasing $T$, the susceptibility to the effects of MR scattering *increases*. As depicted in Fig. S4, even as $\ell_{mc}$ is the shortest length scale, $\ell_{mr}$ has a dominant effect on $R_{nl}$ for $\Delta x \gtrsim 3$ µm ($\ell_{mr} = 14.3$ µm, corresponding to $T = 28$ K). This behavior can be understood by considering the change in current vortex pattern as we fix $\ell_{mr}$ and decrease $\ell_{mc}$. Figure S5 reveals that for fixed $\ell_{mr} = 64.5$ µm the large device-scale current vortex increasingly localizes as $\ell_{mc}$ decreases, before finally disappearing when $\ell_{mc}$ decreases below a threshold. No current vortex or negative $R_{nl}$ appear in the geometry when MC scattering dominates, and the current streamlines then resemble those of diffusive dynamics. Conversely, the hydrodynamic regime is more robust against MR scattering for sufficiently long $\ell_{mc}$ [6].



\end{document}